\documentclass[12pt]{article}
\begin{document}
\begin{center}
{\huge \bf  Quantum Chromodynamics with massive gluons} 
 \\[10mm] 
  S.A. Larin \\ [3mm]
 Institute for Nuclear Research of the
 Russian Academy of Sciences,   \\
 60th October Anniversary Prospect 7a,
 Moscow 117312, Russia
\end{center}

\vspace{15mm}

\begin{abstract}
It is shown that the Lagrangian of Quantum Chromodynamics can be modified
by the adding gluon masses.
On mass-shell renormalizability of the resulting theory is discussed.
\end{abstract}

\newpage

1. The discovery \cite{gvp} of asymptotic freedom in Quantum Chromodynamics
(QCD) has lead to the establishment of QCD as the theory 
of strong interactions. The gauge bosons of the theory, 
the gluons, are considered
to be massless to have gauge invariance and correspondingly 
renormalizability. In the present paper it is shown that the QCD Lagrangian
should be modified by the adding gluon masses to ensure that
QCD does not contradict to experiments. 
On mass-shell
renormalizability of the resulting theory is discussed.

2. The Lagrangian of QCD is
\begin{equation}
\label{lagrangian}
 L_{QCD}  =  -\frac{1}{4}F_{\mu\nu}^a F_{\mu\nu}^a
 +i \overline{\psi}_f\gamma_{\mu}D_\mu \psi_f
- m_f \overline{\psi}_f \psi_f
\end{equation}
\[  
-\frac{1}{\xi}(\partial^\mu A_\mu^a)^2
+\partial^{\mu}\overline{c}^a(\partial_\mu c^a
                 -g f^{abc} c^b A_\mu^c)
+counterterms,
\]
where $F_{\mu\nu}^a=\partial_\mu A_\nu^{a} -\partial_\nu A_\mu^a
 + g f^{abc} A_\mu^b A_\nu^c$ is the gluon field strength tensor,
$D_\mu=\partial_\mu-i g A_\mu^a T^a$ is the covariant derivative.
The quark fields $\psi_f$ transform as 
the fundamental representation of
the colour  group $SU(3)$ ,  $f=u,d,s,c,b,t$ is the flavour index.  
The gluons  $A^a_\mu$ 
transform as the adjoint representation of this group.
$c^a$ are the ghost fields, $\xi$ is the gauge parameter
of the usually chosen general covariant gauge,
$f^{abc}$ are the structure constants of the group,
$T^{a}$ are the generators of the fundamental representation.
$g=g(\mu)$ is the renormalized strong coupling constant, 
$g^2/(16\pi^2)\equiv a_s$,   
$m_f=m_f(\mu)$ is the Lagrangian (renormalized) 
mass of a quark with a flavor $f$, and $\mu$ 
is the renormalization point. The summations over repeated indexes
are assumed.

3. Let us consider the vacuum polarization function 
$\Pi(q^2)$
\begin{equation}
(-q^2g_{\mu\nu}+q_{\mu}q_{\nu})\Pi(q^2) 
= i\int {\rm d} x e^{iqx}
\langle 0| \;
T\;j_{\mu}(x)j_{\nu}(0)\;|0 \rangle.
\end{equation}
where $j_{\mu} =\sum_{f}q_f \overline{\psi}_f \gamma_{\mu}\psi_f $
is the electromagnetic quark current 
and $q_f=2/3,-1/3,...$ is the electromagnetic charge 
of the quark with a flavor $f$.

According to general principles of local quantum field theory the function
$\Pi(q^2)$ satisfies the K\"allen-Lehmann \cite{kl} spectral representation
\begin{equation}
\Pi(q^2) =\frac{1}{12\pi^2}  \int_{4m_{\pi}^2}^\infty \,ds\,\, 
                \frac{R(s)}{s-q^2-i0},
\end{equation}
where the ratio $R(s) = \sigma_{tot}(e^+e^-\rightarrow hadrons)/
\sigma(e^+e^-\rightarrow\mu^+\mu^-) $ is the normalized total
cross-section of electron-positron annihilation into hadrons,
$m_{\pi}$ is a pion mass.

The K\"allen-Lehmann representation determines the analytic properties
of $\Pi(q^2)$ which should be an analytic function in the complex
$q^2$-plane with the cut starting from the first physical threshold, 
i.e. as it is dictated by experiments from the two-pion
threshold $q^2=4m_{\pi}^2$. In particular, one gets for the discontinuity
of $\Pi(q^2)$ on the cut
\begin{equation}
\Delta\Pi(q^2)\equiv\Pi(q^2+i0)-\Pi(q^2-i0)=\left\{
 \begin{array}{l} i~R(q^2)/(6\pi)~~~at~~s>4m_{\pi}^2   \\ 
                        ~~~0~~~~~~~~~~~~~~~at~~s<4m_{\pi}^2.
                 \end{array} \right.
\end{equation}
Perturbative QCD produces the following expression for the discontinuity
\begin{equation}
\Delta\Pi(q^2)_{pQCD} =\theta(q^2)~\rho_{gluon}(q^2)+
 \theta(q^2-4M_u^2)~\rho_{quark}(q^2).
\end{equation}
The gluon spectral density $\rho_{gluon}(q^2)$ contributes 
for $q^2>0$ as it is indicated by the theta-function $\theta(q^2)$.
This is the known zero threshold. It
arises from those absorptive parts of Feynman diagrams
of $\Pi(q^2)$ which are
produced by purely gluonic cuts of the diagrams 
(i.e. by the Cutcosky cuts
which cross only gluon propagators of diagrams). 
As it is well known such
diagrams appear for the first time at the four-loop level 
in the order $a_s^3$ 
(corresponding cuts cross 3 gluon propagators).

The quark spectral density $\rho_{quark}(q^2)$ arises from 
the quark cuts of the diagrams
(i.e. from the cuts which cross two or more quark propagators of
the diagrams). It contributes
for $q^2>4M_u^2$ where
$M_u$ is the perturbative pole mass of the lightest $u$-quark,
defined as the pole of the quark propagator within perturbation
theory. A perturbative quark pole mass 
\begin{equation}
M_f=m_f(\mu) + O(a_s)
\end{equation}
appears after summation of perturbative corrections to a quark
propagator.
It is a renormalization group invariant quantity, i.e.
independent on the renormalization point $\mu$ and on the choice of the
subtraction scheme. In this sense it behaves as a physical object
and that is why it is natural to use this definition of a quark mass
to parametrize the theory.

We will not discuss here the important by themselves questions of convergence
or divergence of corresponding  perturbative QCD series at low or
at high energies. Here we will
just accept that our conventional perturbation theory is adequate
to the exact solution of the theory, i.e. it correctly reproduces
the perturbative expansion of the exact solution. 

Hence one gets within QCD that $\Delta\Pi(q^2)$ 
is non-zero in the energy interval $0<q^2<4m_{\pi}^2$ since the perturbative
contribution $\Delta\Pi(q^2)_{pQCD}$ is non-zero in this interval.
And we would like to stress here that one should get in QCD an exact zero below
the two-pion threshold as it is dictated by experiments.
There are of course also non-perturbative contributions, i.e. contributions
of the type of $e^{-1/a_s}$ which are invisible in the perturbative
expansion at $a_s=0_{+}$
\[e^{-1/a_s}=0\cdot a_s + 0\cdot a_s^2 + ...
\]
At this point
it is interesting to consider examples how non-perturbative contributions
can be separated from perturbative ones. We consider first the following
function
\[
\frac{1}{e^{a_s}+e^{-1/a_s}}
\]
which contains both perturbative and non-perturbative contributions.

Let us nullify in this function the non-perturbative object $e^{-1/a_s}$.
Then we are left with the purely perturbative function $\frac{1}{e^{a_s}}$.
Let us now subtract and add this function to the original one:
\begin{equation}
\frac{1}{e^{a_s}+e^{-1/a_s}}\equiv
\frac{1}{e^{a_s}+e^{-1/a_s}}-\frac{1}{e^{a_s}}+\frac{1}{e^{a_s}}=
-\frac{e^{-1/a_s}}{e^{a_s}+e^{-1/a_s}}+e^{-a_s},
\end{equation}
where  two first terms are summed into one. Thus the original function
is presented as the sum of the purely non-perturbative and purely
perturbative terms.

Let us consider one more simple example:
\[
 \sin\left(e^{a_s}+e^{-1/a_s}\right).
\]
We again nullify in the above function
the non-perturbative object $e^{-1/a_s}$ to get the purely
perturbative function $\sin\left(e^{a_s}\right)$. 
Then we subtract and add this
perturbative functin to the original one:
\begin{equation}
\sin\left(e^{a_s}+e^{-1/a_s}\right)=\left[\sin\left(e^{a_s}+e^{-1/a_s}\right)-
\sin\left(e^{a_s}\right)\right]+\sin\left(e^{a_s}\right).
\end{equation}
Here in the square brackets we have the purely non-perturbative function
and outside - the purely perturbative one.

The general rule is as follows. In an original function which is a mix of
perturbative and non-perturbative contributions one should nullify all
non-perturbative objects getting in this way a purely perturbative function.
Then one should subtract and add this perturbative function to the original
one. The difference of the original function and the purely perturbative one
will form the purely non-perturbative contribution.

Let us now return to the analysis of the spectral dencity. We note
that non-perturbative contributions can not exactly cancel the perturbative
contribution in the continuous interval $0<q^2<4m_{\pi}^2$ because
of the different dependence on $a_s$. To get that 
$\Delta\Pi(q^2)=0$ at $0<q^2<4m_{\pi}^2$ in agreement with experiments
one should move perturbative
gluon and quark thresholds above $q^2=4m_{\pi}^2$.
That is why ome should introduce the non-zero Lagrangian gluon masses.

The first naive objection here is that nobody trusts perturbation theory
below the two-pion threshold, i.e. that the corresponding perturbative series
is heavily divergent in this energy region. But for us
here only the principal existence of the pertubative series
with finite coefficients below the two-pion threshold is of importance
independently on the question of its divergence.

Thus one obtains the following restrictions on the 
(perturbative pole)
masses of gluons and quarks
\begin{equation}
(3M_{gl})^2>4m_{\pi}^2,
\end{equation}
\[ 4M_u^2>4m_{\pi}^2.
\]
Although the restriction on $M_u$ seems to be quite strong for
the lightest u-quark it is not excluded from the first principles.

4. To construct QCD with massive gluons we will follow the approach
of \cite{lar1}. 
Presently this is the only known way to get (on mass-shell) renormalizable
theory of massive gluons without color scalars  (color scalars 
are rejected by experiments). 
Within this approach one starts from a renormalizable
theory with scalar fields using the Englert-Brout-Higgs mechanism 
of spontaneous symmetry breaking \cite{hig}
and after transition to the unitary gauge removes remaining massive scalar
fields. Thus we add to the massless QCD Lagrangian (\ref{lagrangian}) 
the scalar part to begin with the following general Lagrangian
\begin{equation}
L_{QCD+scalars}=
 -\frac{1}{4}F_{\mu\nu}^a F_{\mu\nu}^a
 +i \overline{\psi}_f\gamma_{\mu}D_\mu \psi_f
- m_f \overline{\psi}_f \psi_f +
\end{equation}
\[
\left(D_\mu\Phi\right)^+D_\mu\Phi +\left(D_\mu\Sigma\right)^+D_\mu\Sigma
-\lambda_1\left(\Phi^+\Phi -v_1^2\right)^2 
-\lambda_2\left(\Sigma^+\Sigma-v_2^2\right)^2
\]
\[
-\lambda_3\left(\Phi^+\Phi+\Sigma^+\Sigma-v_1^2-v_2^2\right)^2
-\lambda_4\left(\Phi^+\Sigma\right)\left(\Sigma^+\Phi\right)
\]
\[
+L_{gf}+ L_{gc}+counterterms,
\]
where we introduced two triplets $\Phi(x)$ and $\Sigma(x)$ 
of complex scalar fields 
in the fundamental representation of the $SU(3)$ color
group to get all gluon massive.
$L_{gf}$ is the gauge fixing part of the Lagrangian in some chosen gauge
and $L_{gc}$ is the corresponding gauge compensating part with the 
Faddeev-Popov ghost fields.

We can choose the following shifts of scalar fields by the quantities
$v_1$ and $v_2$
to generate masses of all eight gluons
\begin{equation}
\Phi(x) =\left( \begin{array}{l} \phi_{1}(x)+ i\phi_{2}(x)+v_1   \\ 
                \phi_{3}(x)+i\phi_4(x)\\
                \phi_5(x)+i\phi_6(x) \end{array} \right),~~~
\Sigma(x) =\left( \begin{array}{l} \sigma_{1}(x)+ i\sigma_{2}(x)   \\ 
                \sigma_{3}(x)+i\sigma_4(x)+v_2\\
                \sigma_5(x)+i\sigma_6(x) \end{array} \right).
\end{equation}
Choosing for simplicity $v_1=v_2\equiv v$ one obtains the following
massive terms for gluons in the Lagrangian
\begin{equation}
\label{gmasses}
L_M=M^2\left[(A^1)^2+(A^2)^2+(A^3)^2+\frac{1}{2}(A^4)^2+\right.
\end{equation}
\[ \left.
\frac{1}{2}(A^5)^2+\frac{1}{2}(A^6)^2+\frac{1}{2}(A^7)^2+
\frac{1}{3}(A^8)^2\right],
\]
where $M^2 \equiv g^2 v^2$ is the gluon mass
parameter of the theory.

After the chosen shifts the following four combinations of scalar fields
\begin{equation} 
\phi_1+\frac{\lambda_3}{\lambda_1+\lambda_3}\sigma_3,~~~\sigma_3,~~~
\sigma_1+\phi_3,~~~\sigma_2-\phi_4
\end{equation}
become massive Higgs particles.

The following eight combinations
\begin{equation} 
\sigma_1-\phi_3,~~\phi_4+\sigma_2,~~\phi_2-\sigma_4,~~\phi_2+\sigma_4,~~
\phi_5,~~\phi_6,~~\sigma_5,~~\sigma_6
\end{equation}
become massless Goldstone ghosts.

Now one can make transition to the unitary gauge. All ghost fields
as usual disappear from the Lagrangian.
Following the approach of \cite{lar1} one can remove
in the unitary gauge all Higgs fields from the Lagrangian preserving
on mass-shell renormalizability of the theory. 
The Lagrangian of the resulting theory is
\begin{equation}
L_{massive~QCD}=L_M
 -\frac{1}{4}F_{\mu\nu}^a F_{\mu\nu}^a
 +i \overline{\psi}_f\gamma_{\mu}D_\mu \psi_f
- m_f \overline{\psi}_f \psi_f+counterterms,
\end{equation}
where $L_M$ is given in eq.(\ref{gmasses}).

Let us note that
on mass-shell renormalizability does not mean that one should
consider quarks and gluons as free external particles.
It means that in the $SU(3)\times SU(2)\times U(1)$ theory
the $S$-matrix elements with the physical external particles
will be finite.

One can calculate the one-loop 
$\beta$-function in this theory to obtain
for a massless renormalization scheme 
(i.e. a scheme where renormalization group functions do not
depend on masses) the following result
\begin{equation}
\beta(a_s)=\mu^2\frac{\partial a_s}{\partial \mu^2}=
\sum_{i\ge 0}\beta_i a_s^{i+2},
\end{equation}
\[
\beta_0=-\frac{7}{2}C_A+\frac{4}{3}T_F n_f,
\]
here $C_A=3$ is the Casimir operator of the adjoint representation
of the $SU(3)$ color group, $T_F=1/2$ is the trace normalization
of the fundamental representation, $n_f$ is the number of active quark flavors.

Thus asymptotic freedom remains valid in the considered theory with massive
gluons.

  {\bf Acknowledgments.}

The author is grateful to collaborators of the Theory division of INR
for helpful discussions. The work is supported in part by the grant
for the Leading Scientific Schools NS-5590.2012.2.

\end{document}